\documentclass[preprint,12pt]{elsarticle}
\usepackage{epsfig}
\usepackage{rotating}
\usepackage[usenames]{color}
\usepackage{multirow}

\begin{document}
\begin{frontmatter}

\title{Improvement of radiopurity level of enriched $^{116}$CdWO$_4$ and ZnWO$_4$ crystal scintillators
by recrystallization}

\author[itep]{A.S.~Barabash}
\author[infn-tv]{P.~Belli}
\author[infn-tv,univ-tv]{R.~Bernabei\corref{cor1}}
\cortext[cor1]{Corresponding author}
 \ead{rita.bernabei@roma2.infn.it}
\author[niic]{Yu.A.~Borovlev }
\author[lngs]{F.~Cappella}
\author[lngs]{V.~Caracciolo}
\author[lngs]{R.~Cerulli}
\author[kinr]{F.A.~Danevich}
\author[infn-ls,univ-ls]{A.~Incicchitti}
\author[kinr]{V.V.~Kobychev}
\author[itep]{S.I.~Konovalov}
\author[lngs]{M.~Laubenstein}
\author[kinr,infn-ls]{V.M.~Mokina}
\author[kinr]{O.G.~Polischuk}
\author[sigm]{O.E.~Safonova}
\author[niic]{V.N.~Shlegel}
\author[kinr,infn-ls]{V.I.~Tretyak}
\author[isma]{I.A.~Tupitsyna}
\author[itep]{V.I.~Umatov}
\author[cml]{V.N.~Zhdankov}

\address[itep]{National Research Centre ``Kurchatov Institute'', Institute of Theoretical and Experimental Physics, 117218 Moscow, Russia}
\address[infn-tv]{ INFN, sezione di Roma ``Tor Vergata'', I-00133 Rome, Italy}
\address[univ-tv]{ Dipartimento di Fisica, Universit\`a di Roma ``Tor Vergata'', I-00133 Rome, Italy}
\address[niic]{Nikolaev Institute of Inorganic Chemistry, 630090 Novosibirsk, Russia}
\address[lngs]{INFN, Laboratori Nazionali del Gran Sasso, I-67100 Assergi (AQ), Italy}
\address[kinr]{Institute for Nuclear Research, MSP 03680 Kyiv, Ukraine}
\address[infn-ls]{INFN, sezione di Roma, I-00185 Rome, Italy}
\address[univ-ls]{Dipartimento di Fisica, Universit\`a di Roma ``La Sapienza'', I-00185 Rome, Italy}
\address[sigm]{V.S. Sobolev Institute of Geology and Mineralogy of
the Siberian Branch of the RAS, 630090 Novosibirsk, Russia}
\address[isma]{Institute of Scintillation Materials, 61001 Kharkiv, Ukraine}
\address[cml]{CML Ltd., 630090 Novosibirsk, Russia}

\begin{abstract}
As low as possible radioactive contamination of a detector plays a
crucial role to improve  sensitivity of a double beta decay
experiment. The radioactive contamination of a sample of
$^{116}$CdWO$_4$ crystal scintillator by thorium was reduced by a
factor $\approx10$, down to the level 0.01 mBq/kg ($^{228}$Th), by
exploiting the recrystallization procedure. The total alpha
activity of uranium and thorium daughters was reduced by a factor
$\approx3$, down to 1.6 mBq/kg. No change in the specific activity
(the total $\alpha$ activity and $^{228}$Th) was observed in a
sample of ZnWO$_4$ crystal produced by recrystallization after
removing $\approx 0.4$ mm surface layer of the crystal.
\end{abstract}


\end{frontmatter}

\section{Introduction}

\begin{figure}[htb]
 \begin{center}
 \mbox{\epsfig{figure=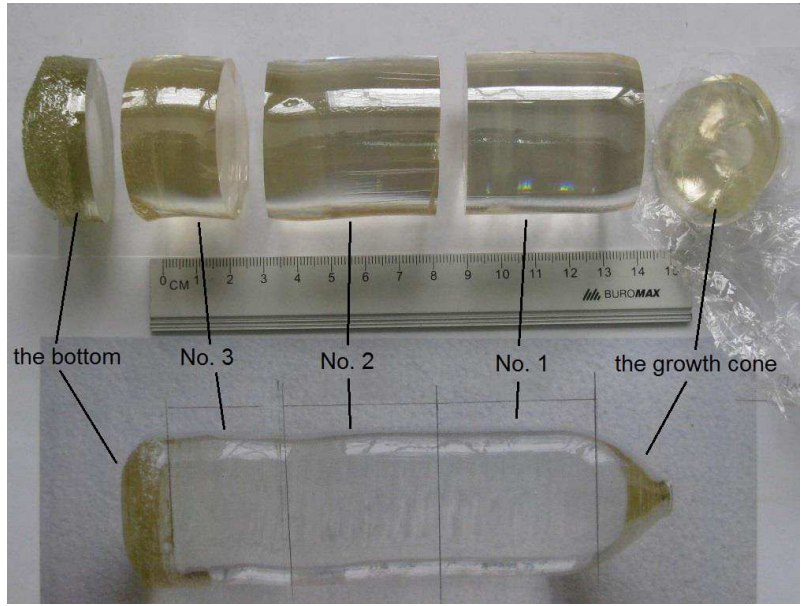,height=8.0cm}}
 \caption{(Color online) Boule of enriched $^{116}$CdWO$_4$ crystal and the crystal samples produced for the double beta decay experiment.}
 \label{fig:samples}
 \end{center}
 \end{figure}

High quality large volume cadmium tungstate crystal scintillators
were developed from cadmium enriched in $^{116}$Cd to 82\%
($^{116}$CdWO$_4$) for the Aurora experiment to investigate the
double beta ($2\beta$) decay of $^{116}$Cd \cite{Barabash:2011}.
Three crystal scintillators (586 g, No. 1; 589 g, No. 2; 326 g,
No. 3) have been cut from the crystal boule (see Fig.
\ref{fig:samples}). The experiment is in progress with two
scintillators (No. 1 and No. 2) installed in the DAMA/R\&D low
background set-up at the Gran Sasso underground laboratory (LNGS) of the
INFN (Italy) (average depth $\approx 3650$ m w.e.
\cite{Ambrosio:1995}). Preliminary results of the experiment were
reported in conference proceedings
\cite{Poda:2014,Polischuk:2015,Danevich:2016}.

The radiopurity level of the crystal scintillators plays a crucial
role in double beta decay experiments. The radioactive
contamination of the crystal scintillators No. 1 and No. 2 was
estimated from the experimental data of the Aurora experiment
\cite{Barabash:2011,Poda:2013,Danevich:2013}, while the
radioactive contamination of the crystal sample No. 3 was measured
by HPGe detector \cite{Barabash:2011} and also in the present work
by scintillation counting method (see Sections \ref{LBmsr} and
\ref{results}). The contamination of the rest of the melt
remaining in the platinum crucible after the crystal growth was
measured with the help of the GePaolo ultra-low background HPGe
$\gamma$ spectrometer of the STELLA (SubTErranean Low Level Assay)
facility at LNGS \cite{Laubenstein:2004}. The radioactive
contamination of the crystal samples and of the rest after the
crystal growth is presented in Table \ref{table:act}. The
contamination of the rest material by potassium, radium and
thorium is substantially higher than the contamination of the
crystals. An increase of radioactive contamination by thorium
along the crystal boule from the growth cone to the bottom of the
boule was observed too. The observed variation of the thorium
specific activity, and a much higher contamination of the rest
after the crystal growing process, can be explained by segregation
of the radioactive elements in the CdWO$_4$ crystal growth
process, which provides a possibility to improve the radiopurity
level of the $^{116}$CdWO$_4$ crystal scintillators by
re-crystallization. It should be stressed that multiple
crystallization is well-known approach to improve optical quality
(see, e.g., \cite{Kobayashi:1997}) and radiopurity level
\cite{Danevich:2011,Armengaud:2016} of crystal scintillators.
Thus, the $^{116}$CdWO$_4$ crystal sample No. 3 was recrystallized
to study the impact of the second crystallization on the
radioactive contamination of cadmium tungstate crystals.

\nopagebreak

\begin{table}[h!]
\caption{Specific activities of radioactive contaminants in the
different parts of the $^{116}$CdWO$_4$ crystals produced from the
boule, and in the rest of the melt after the crystal growth. The
limit on activity of $^{228}$Ra in the sample No.~3 was estimated
by analysis of $^{228}$Ac $\gamma$ quanta in the data accumulated
by the HPGe detector \cite{Barabash:2011}. Reference date is May
2014.}
\begin{center}
\begin{tabular}{|l|l|l|l|l|l|}
\hline
$^{116}$CdWO$_4$ sample,      & \multicolumn{4}{c|}{Radioactive contamination (mBq/kg)}   & Method  \\
 \cline{2-5}
 mass (see also Fig. \ref{fig:samples})  & $^{40}$K  & $^{226}$Ra    & $^{228}$Ra    & $^{228}$Th            & ~ \\
 \hline
 No. 1, 586 g                       & $\leq0.9$ & $\leq0.005$   & ~             & 0.02(1)               & Scintillation counting \\
 No. 2, 589 g                       & $\leq0.9$ & $\leq0.005$   & ~             & 0.04(1)               & Scintillation counting  \\
 No. 3, 326 g                       & $\leq1.2$ & $\leq0.1$     & $\leq1.3$     & 0.10(1)               & Scintillation counting \\
 Rest of the melt                   & ~         & ~             & ~             & ~                     & ~ \\
 after the crystal                  & ~         & ~             & ~             & ~                     & ~ \\
 growth, 264 g                      & 27(11)    & 64(4)         & 9(2)          & 10(2)                 & HPGe $\gamma$ spectrometry  \\
 \hline

\end{tabular}
\end{center}
\label{table:act}
\end{table}

As regards the zinc tungstate (ZnWO$_4$) crystal scintillators,
they are promising detectors to search for double beta decay
\cite{Danevich:2005,Belli:2011a}, dark matter
\cite{Bavykina:2008,Kraus:2009,Cappella:2013}, eka-elements
\cite{Belli:2015}, and investigation of rare alpha decays
\cite{Casali:2016}. The detector has a very low level of
radioactive contamination \cite{Belli:2011b}. The effect of the
recrystallization procedure on the radioactive contamination of
the ZnWO$_4$ crystal was also investigated in Ref.
\cite{Belli:2011b}. Although one could expect an improvement of
the material purity level after a second crystallization, the
radioactive contamination of the sample obtained by
recrystallization was even slightly higher, on the contrary of the
case of CdWO$_4$. This could be explained by a concentration of
radioactive impurities in a thin surface layer of the ZnWO$_4$
crystal\footnote{Indeed, the ratio surface/volume is higher in the
smaller crystal sample, which could explain the higher value of
the specific total $\alpha$ activity in the ZnWO$_4$ crystal after
recrystallization.}. In Ref. \cite{Danevich:1995} a substantial
reduction of cadmium tungstate crystal scintillators radioactive
contamination was achieved by removing a thin ($0.8-1.5$ mm)
surface layer of the crystals. In the present work a possible
concentration of radioactive impurities in a surface layer of the
ZnWO$_4$ crystal obtained by recrystallization was also checked.

\section{Samples of $^{116}$CdWO$_4$ and ZnWO$_4$ crystal scintillators}
\label{samples}

\begin{figure}[htbp]
\begin{center}
\resizebox{0.463\textwidth}{!}{\includegraphics{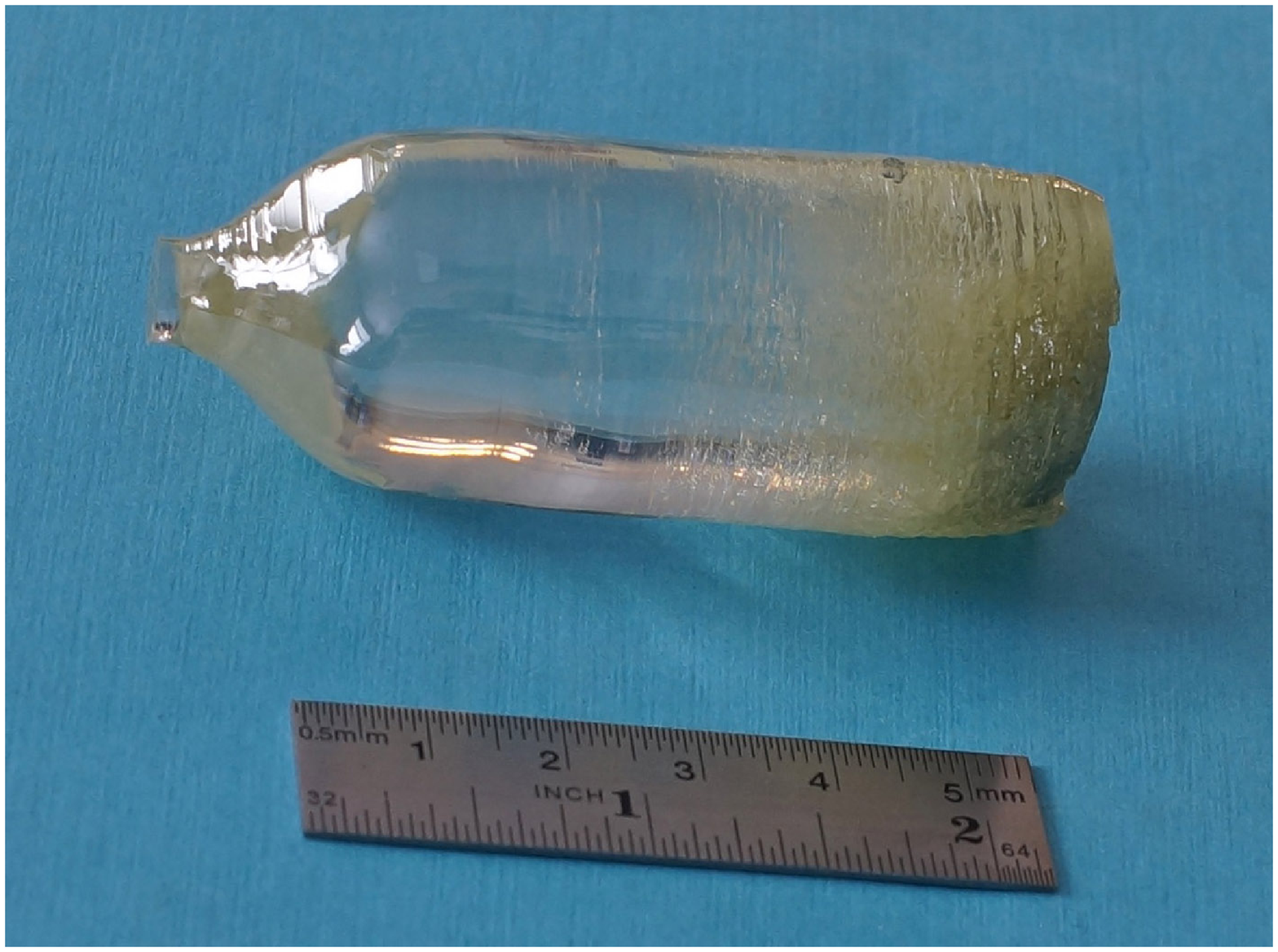}}
\resizebox{0.49\textwidth}{!}{\includegraphics{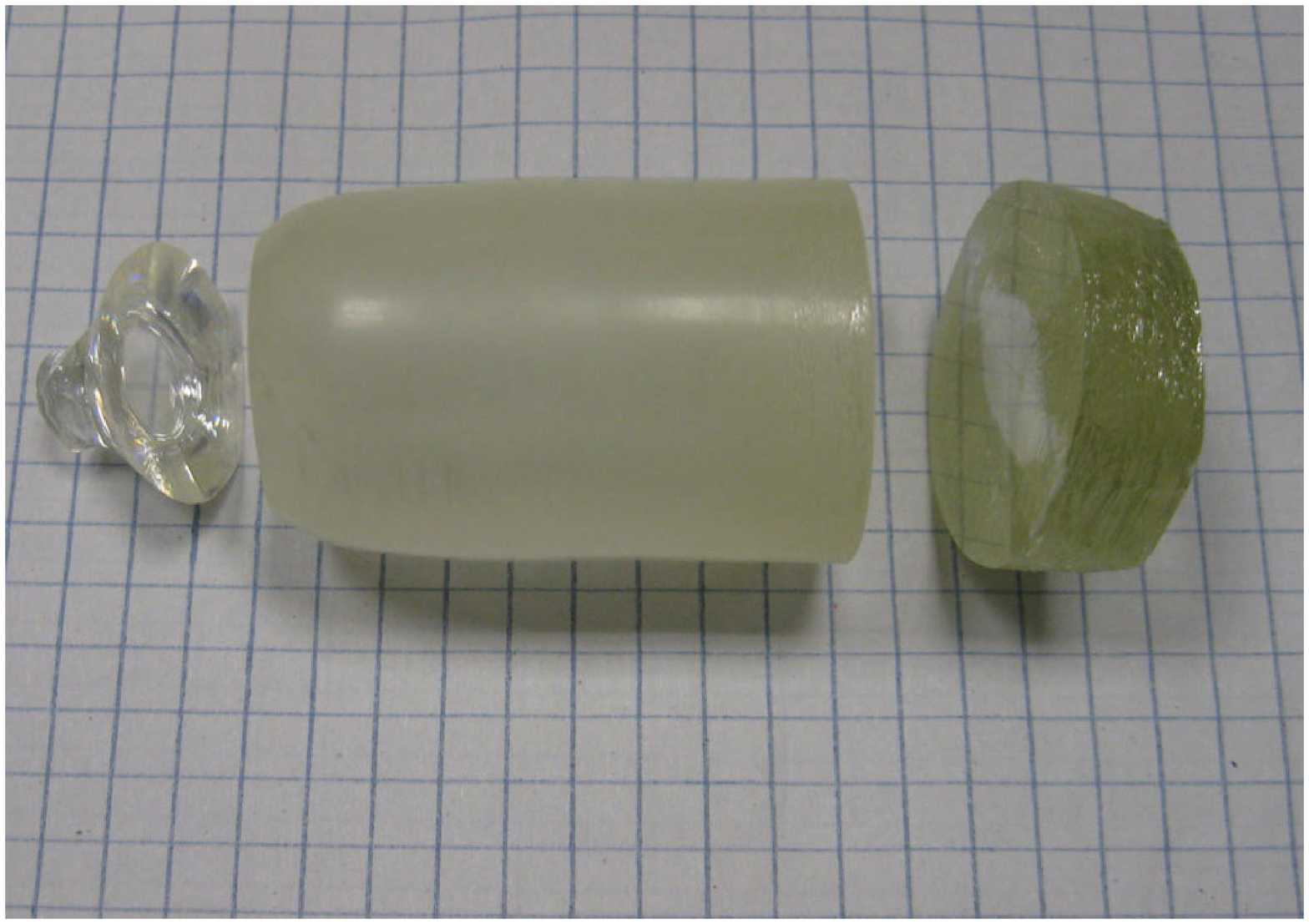}}
\caption{(Color online) Left: Boule of enriched $^{116}$CdWO$_4$
crystal obtained by recrystallization. The conic part of the boule
is the beginning of the crystal growth. Right: The boule cut into
three parts. The side surface of the central part (used in the
present study) is diffused to improve the light collection.}
\end{center}
 \label{fig:cryst}
\end{figure}

The sample No. 3 of the $^{116}$CdWO$_4$ crystal was
recrystallized by the low-thermal-gradient Czochralski technique
\cite{Pavlyuk:1992,Borovlev:2001,Galashov:2009} in a platinum
crucible 50 mm in diameter with 99.93\% Pt purity level.
The crystallization rate was 1.5 mm/h. A crystal boule with mass
286 g (see Fig. \ref{fig:cryst}, left) was grown, which is 88\%
of the sample before recrystallization. The boule was cut into three parts
as shown in Fig. \ref{fig:cryst}, right. The central part with
mass 195 g (60\% of the sample before recrystallization) was used to test
the radioactive contamination. The side surface of the sample was
made opaque by grinding paper to improve the scintillation light
collection.

\nopagebreak

On the other hand, a ZnWO$_4$ crystal sample with mass 141 g was
obtained by recrystallization -- with ordinary Czochralski method
-- of a 699 g ZnWO$_4$ crystal\footnote{The low yield of ZnWO$_4$
crystal (20\%) is typical for ordinary Czochralski method, in
contrast to the low-thermal-gradient Czochralski technique (used
for the $^{116}$CdWO$_4$ re-crystallization) which allows to
obtain crystal boules with the mass of up to $\sim90\%$ of the
initial charge \cite{Grigoriev:2014}.}. Both crystals were used in
the experiments to search for double beta decay of zinc and
tungsten \cite{Belli:2011a}. The radioactive contamination of the
samples is summarized in Ref. \cite{Belli:2011b}. The total alpha
activity in the recrystallized sample, 0.47(7) mBq/kg, is even
higher than that of the initial crystal, 0.18(3) mBq/kg. Such a
difference in the specific activity could be explained by
concentration of uranium, thorium and their daughters in the
surface layer of the crystals. The side surface layer $\approx0.4$
mm of the recrystallized sample was removed by a diamond-needle
file to test the assumption that the impurities were concentrated
in a thin surface layer. Then the side surface of the crystal was
made opaque by grinding paper. The mass of the crystal after the
treatment was 133 g.

All the crystal samples were carefully cleaned using special low
radioactive detergents, ultra-pure nitric acid and water before
the low background measurements described in the next section.

\section{Low background measurements}
\label{LBmsr}

The radioactive contamination of the samples was measured in the
\newline{DAMA/CRYS} set-up at the Gran Sasso underground
laboratory. This is a low background facility realized for low
counting experiments using scintillation detectors. The passive
shield of the set-up is made of high purity copper (11 cm), lead
(10 cm), cadmium (2 mm), and polyethylene (10 cm). The set-up is
sealed and continuously flushed by high purity nitrogen gas to
prevent the detector to be in contact with the environmental air.
The inner volume available for scintillation detector(s)
installation is ($82\times 16\times 14$) cm$^3$.

The $^{116}$CdWO$_4$ crystal scintillator before the
recrystallization and the ZnWO$_4$ crystal were viewed by a
low-radioactive 3 inch green enhanced photomultiplier (PMT,
Hamamatsu R6233MOD) through a high purity quartz light guide
$\oslash 7.6\times 10$ cm.

The $^{116}$CdWO$_4$ crystal scintillator after the
recrystallization was installed inside a cavity ($\oslash 47
\times 59$ mm) in the central part of a polystyrene light guide
$\oslash66\times 312$ mm. The cavity was filled with high-purity
silicon oil. Two high-purity quartz light guides $\oslash 66\times
100$ mm were optically connected to the opposite sides of the
polystyrene light guide. The scintillation light was collected by
two low-radioactive 3 inch PMTs (EMI9265–B53/FL).

An event-by-event data acquisition system based on a four channels
14 bit transient digitizer (CAEN DT5724D) records the time of each
event and the pulse profile with a sampling rate 100 MS/s.

The energy scale and energy resolution of the $^{116}$CdWO$_4$ and
ZnWO$_4$ detectors were measured with $^{22}$Na, $^{60}$Co,
$^{137}$Cs, $^{133}$Ba, and $^{228}$Th $\gamma$ sources.

\section{Data analysis, results and discussion}
\label{results}

\subsection{Pulse-shape discrimination}

\begin{figure}[htb]
 \begin{center}
 \mbox{\epsfig{figure=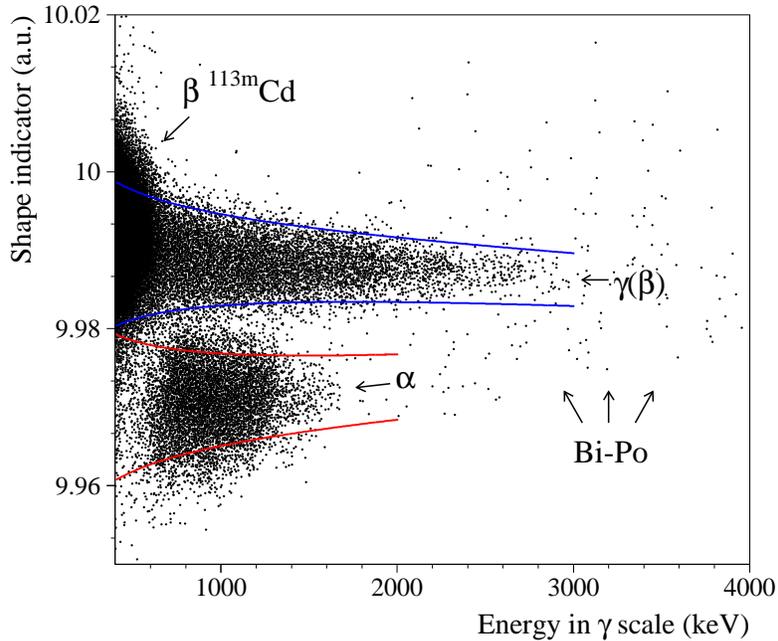,height=8.5cm}}
 \caption{(Color online) Shape indicator (see text) versus the energy
accumulated over 2394 h in the low-background set-up with the
$^{116}$CdWO$_4$ crystal scintillator before the
recrystallization. The two sigma intervals for the shape indicator
values corresponding to $\gamma$ quanta ($\beta$ particles) and
$\alpha$ particles are shown. The $\alpha$ events are clearly
separated from $\gamma$ and $\beta$ events. The counts distributed
in the energy region $2-4$ MeV outside of the $\alpha$ and
$\gamma(\beta)$ belts are events of Bi$-$Po decays.}
 \label{fig:psd}
 \end{center}
 \end{figure}

The $\alpha$ events in the CdWO$_4$ and ZnWO$_4$ detectors were
selected with the help of the pulse-shape discrimination procedure
based on the optimal filter method proposed by E. Gatti and F. De
Martini \cite{Gatti:1962}. For each signal $f(t)$, the numerical
characteristic of its shape (shape indicator, $SI$) was defined
as: $SI = \sum f(t_k) \times P(t_k)/\sum f(t_k)$. The sum is over
the time channels $k$, starting from the origin of signal and up
to 30 $\mu$s, and $f(t_k)$ is the digitized amplitude (at the time
$t_k$) of a given signal. The weight function $P(t)$ was defined
as:
$P(t)=[f_{\alpha}(t)-f_{\gamma}(t)]/[f_{\alpha}(t)+f_{\gamma}(t)]$,
where $f_{\alpha}(t)$ and $f_{\gamma}(t)$ are the reference pulse
shapes for $\alpha$ particles and $\gamma$ quanta. By using this
approach, $\alpha$ events were clearly separated from $\gamma$
($\beta$) events as shown in Fig. \ref{fig:psd} where the scatter
plot of the shape indicator versus energy is depicted for the data
of the low background measurements over 2394 h with the
$^{116}$CdWO$_4$ detector before the recrystallization.
The events with energy 2--4 MeV outside the $\gamma$ ($\beta$) and
$\alpha$ belts are caused by the fast decay chains (Bi-Po events):
$^{212}$Bi$-^{212}$Po (daughters of $^{232}$Th) and
$^{214}$Bi$-^{214}$Po ($^{238}$U).

 \nopagebreak

\begin{figure}[htb]
 \begin{center}
 \mbox{\epsfig{figure=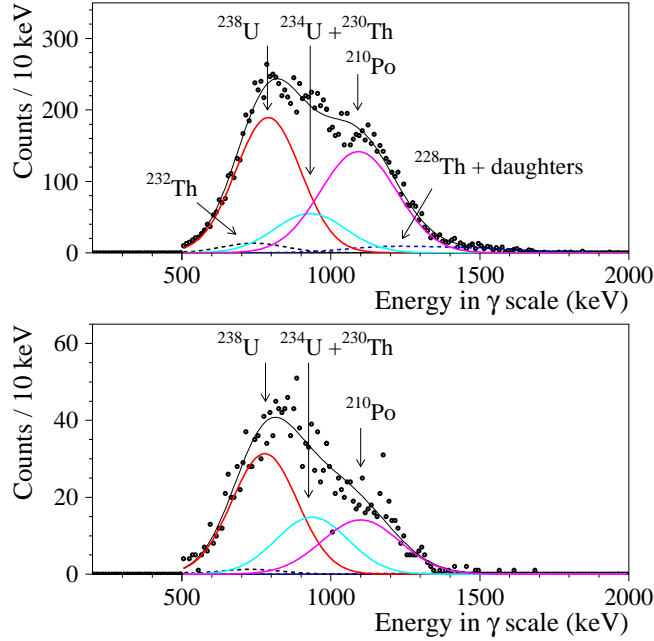,height=8.5cm}}
\caption{(Color online) The energy spectrum of $\alpha$ events
selected by the pulse-shape discrimination from the data
accumulated  in the low-background set-up with the
$^{116}$CdWO$_4$ crystal scintillator before (upper figure, time
of measurement 2394 h) and after the recrystallization (lower
figure, 1623 h).}
 \label{fig:alpha}
 \end{center}
 \end{figure}

The energy spectra of $\alpha$ particles selected by the optimal
filter method from the data accumulated with the $^{116}$CdWO$_4$
crystal scintillators before and after the recrystallization are
presented in Fig. \ref{fig:alpha}. The spectra were fitted by the
model, which includes $\alpha$ peaks of $^{232}$Th, $^{238}$U and
their daughters. Equilibrium in the $^{232}$Th and $^{238}$U
chains was assumed to be broken. The results of the fits are
presented in Fig. \ref{fig:alpha} and in Table \ref{table:116CWO}.
One can see a clear decrease of the total alpha activity in the
$^{116}$CdWO$_4$ crystal scintillator after the recrystallization
by a factor $\approx 3$. Segregation of uranium and lead in the
$^{116}$CdWO$_4$ crystal growth process was also observed in work
\cite{Barabash:2016}, where a sample produced from the growth cone
was tested as low temperature scintillating bolometer.

 \nopagebreak

 \begin{table}[h]
\caption{Radioactive contamination of the $^{116}$CdWO$_4$ crystal
before and after recrystallization.}
\begin{center}
\begin{tabular}{|l|l|l|l|}

\hline
 Chain      & Nuclide       & \multicolumn{2}{|l|}{Activity (mBq/kg)} \\
\cline{3-4}
 ~          & (sub-chain)   & before recrystallization      & after recrystallization   \\
\hline
 $^{232}$Th & $^{232}$Th    & $0.13(7)$                     & $0.03(2)$               \\
 ~          & $^{228}$Th    & 0.10(1)                      & $0.010(3)$               \\
 $^{238}$U  & $^{238}$U     & $1.8(2)$                      & $0.8(2)$             \\
 ~          & $^{226}$Ra    & $\leq0.1$                    & $\leq 0.015$               \\
 ~          & $^{234}$U$+^{230}$Th & $0.6(2)$              & $0.4(1)$                \\
 ~          & $^{210}$Po    & $1.6(2)$                      & $0.4(1)$                  \\
 \hline

 \multicolumn{2}{|l|}{Total $\alpha$} & 4.44(4)              & $1.62(4)$                 \\
 \hline
\end{tabular}
\end{center}
\label{table:116CWO}
\end{table}

\subsection{Time-amplitude analysis to estimate $^{228}$Th activity}
\label{t-A}

The activity of $^{228}$Th in the crystals was estimated with the
help of the time-amplitude analysis (see e.g. Ref.
\cite{Danevich:2001}). The arrival time, the energy and the pulse
shape of each event were used to select the fast decay chain in
the $^{228}$Th sub-chain of the $^{232}$Th family: $^{224}$Ra
($Q_{\alpha} = 5.789$ MeV, $T_{1/2}=3.66$ d) $\rightarrow$
$^{220}$Rn ($Q_{\alpha}=6.405$ MeV, $T_{1/2}=55.6$ s)
$\rightarrow$ $^{216}$Po ($Q_{\alpha} = 6.906$ MeV, $T_{1/2} =
0.145$ s) $\rightarrow$ $^{212}$Pb. As a first step, the $\alpha$
events within an energy interval $0.8-1.85$ MeV were used as
triggers, while for the second events a time interval $0.001-1.0$
s and the energy interval $0.8-1.95$ MeV were required. The
$\alpha$ events were selected in the $\pm2.327\sigma$ "alpha"
region. The energy spectra of $^{220}$Rn and $^{216}$Po $\alpha$
particles, and the distribution of the time intervals between the
events were selected at this step.

As a next step, all the selected pairs ($^{220}$Rn$~-^{216}$Po)
were used as triggers in order to find the events of the decay of
the parent $\alpha$ active $^{224}$Ra. A $1-112$ s time interval
(74.01\% of $^{220}$Rn decays) was chosen to select $\alpha$
events in the energy interval $0.6-1.6$ MeV. Here we took into
account the quenching in the CdWO$_4$ and ZnWO$_4$ scintillators
for $\alpha$ particles \cite{Tretyak:2014} to choose the energy
intervals where the $\alpha$ events of $^{224}$Ra, $^{220}$Rn, and
$^{216}$Po are expected. The energy spectrum of $^{224}$Ra
$\alpha$ particles and the distribution of the time intervals
between $^{224}$Ra and $^{220}$Rn decays were obtained at the
second step of the time-amplitude analysis.

\begin{figure}[htb]
 \begin{center}
 \mbox{\epsfig{figure=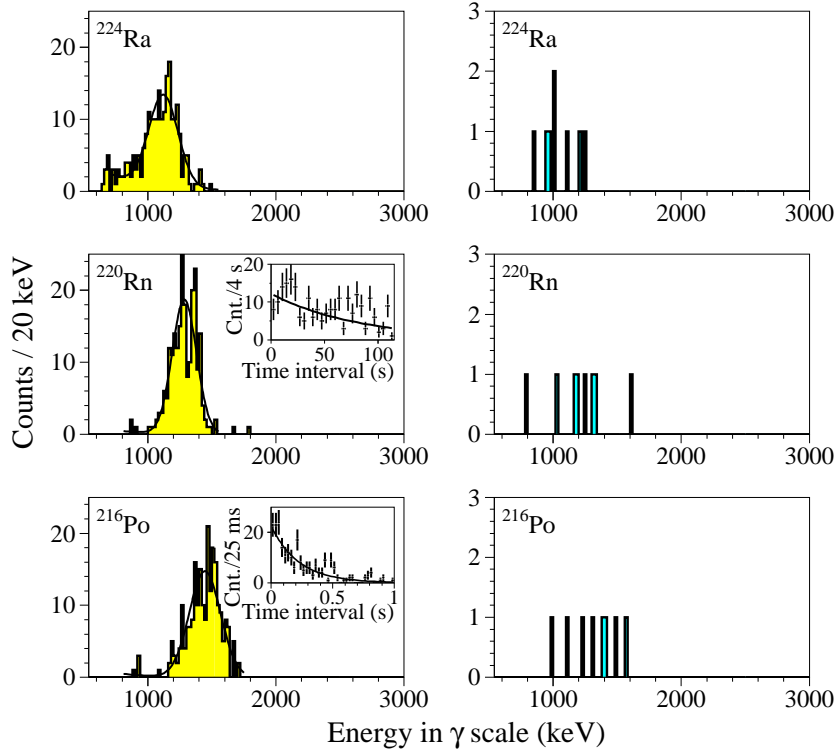,height=10.0cm}}
 \caption{(Color online) Alpha peaks of $^{224}$Ra, $^{220}$Rn and
$^{216}$Po selected by the time-amplitude analysis from the data
accumulated during 2394 h with the $^{116}$CdWO$_4$ scintillator
before the recrystallization (left column of figures) and over
1623 h with the $^{116}$CdWO$_4$ crystal after the
recrystallization (right column). (Insets) The obtained half-lives
of $^{220}$Rn ($77^{+35}_{-18}$ s) and $^{216}$Po
($0.14^{+0.04}_{-0.03}$ s) are in agreement with the table values
(55.6 s and 0.145 s, respectively).}
 \label{fig:t-a}
 \end{center}
 \end{figure}

The $\alpha$ peaks and the distributions of the time intervals
between events obtained by the time-amplitude analysis of the data
measured with the $^{116}$CdWO$_4$ crystal scintillator before the
recrystallization (see left column in Fig. \ref{fig:t-a}) are in
agreement with those expected for the chain. The activity of
$^{228}$Th in the crystal before the recrystallization was
calculated as 0.10(1) mBq/kg. The results of the time-amplitude
analysis of the data accumulated with the $^{116}$CdWO$_4$
detector after the recrystallization are presented in Fig.
\ref{fig:t-a} (right). The specific activity of $^{228}$Th after
recrystallization decreased by $\approx 10$ times to the level of
0.010(3) mBq/kg, which is the result of the strong segregation of
thorium in the CdWO$_4$ crystal growth process. It should be
stressed that thorium is the most harmful radioactive
contamination for double beta decay experiments due to a large
energy release ($Q_{\beta}=4999$ keV) in the beta decay of
$^{208}$Tl (daughter of $^{228}$Th from the $^{232}$Th family).

The summary of radioactive contamination of the $^{116}$CdWO$_4$
crystals before and after the recrystallization process (or limits
on their activities) is given in Table \ref{table:116CWO}.

 \nopagebreak

\subsection{Radioactive contamination of ZnWO$_4$ crystal scintillator with removed surface
layer}

The total $\alpha$ activity in the ZnWO$_4$ crystal after the
surface layer treatment was estimated with the help of the
pulse-shape discrimination of the data accumulated over 1445 hours
to be 0.45(3) mBq/kg, i.e. the same as before the surface treatment:
0.47(7) mBq/kg. The activity of $^{228}$Th was estimated by using
the time-amplitude selection of the events as $\leq 0.003$ mBq/kg
(only one $^{224}$Ra$~-^{220}$Rn$~-^{216}$Po event was selected),
while a specific activity 0.002(1) mBq/kg was measured in the
ZnWO$_4$ crystal before the surface layer treatment.

Therefore, no reduction of the specific activity of $^{228}$Th and
of the total $\alpha$ activity of $^{232}$Th and $^{238}$U and their
daughters was observed after removing the 0.4 mm surface layer.
Taking into account the rather similar chemical and physical
properties of ZnWO$_4$ and CdWO$_4$ crystals, one can assume that
the concentration of uranium and thorium in a thin (a few tenths
of a millimeter) surface layer of the enriched $^{116}$CdWO$_4$
crystal scintillators reported in Ref. \cite{Danevich:1995} can be
explained by a pollution of the $^{116}$CdWO$_4$ crystal due to
diffusion of radioactive elements in the crystal during the
annealing process. In fact, in such an annealing process the
$^{116}$CdWO$_4$ crystal boule was in
direct contact with the ceramic oven at high temperature during
approximately one day. An R\&D to test this assumption is in
progress aiming the development of highly radiopure cadmium and
zinc tungstate crystal scintillators for low counting experiments.

\section{Conclusions}

The recrystallization of a sample of cadmium tungstate crystal
enriched in $^{116}$Cd reduced the thorium contamination of the
sample by one order of magnitude to the level of 0.01 mBq/kg
($^{228}$Th). The total alpha activity due to uranium, thorium,
and their daughters decreased by a factor $\approx 3$ to the level
of 1.6 mBq/kg. This finding, as well as the much higher
contamination of the rest of the melt remaining in the platinum
crucible after the crystal growing process by potassium, radium
and thorium, indicates strong segregation of the radioactive
elements in the CdWO$_4$ crystals growing process. This feature
can be used in order to produce highly radiopure CdWO$_4$ crystal
scintillators for high sensitivity double beta decay experiments.

Moreover, possible concentration of radioactive elements (the activity of
$^{228}$Th and the total $\alpha$ activity of uranium, thorium and
their daughters) in a thin, $\approx 0.4$ mm, surface layer was
not observed in the ZnWO$_4$ crystal. Therefore the concentration
of $\alpha$ active uranium and thorium and their daughters in a thin
surface layer of the enriched $^{116}$CdWO$_4$ crystal
scintillators reported in Ref. \cite{Danevich:1995} was not due to the
growing process but most likely due to the radioactive elements
diffusion into the $^{116}$CdWO$_4$ crystal during the annealing
process in a ceramic oven.

\section{Acknowledgement}

F.A.D. acknowledges the support of the Tor Vergata University
(Rome, Italy).

\end{document}